# Skyrmion Emergence via Domain Wall Anchoring through Vertical Bloch Line

**Authors:** Suyeong Jeong[1*], Dae-Han Jung[1], Hee-Sung Han[2], Ganghwi Kim[1], Myeonghwan Kang[1], Mi-Young Im[4], Younggun Park[3] and Ki-Suk Lee[1,3*]

**Affiliations:**

[1]Department of Materials Science and Engineering, Ulsan National Institute of Science and Technology (UNIST), Ulsan 44919, Republic of Korea

[2]Department of Materials Science and Engineering, Korea National University of Transportation, Chungju 27469, Republic of Korea

[3]Graduate School of Semiconductor Materials and Devices Engineering, Ulsan National Institute of Science and Technology (UNIST), Ulsan 44919, Republic of Korea

[4]Materials Sciences Division, Lawrence Berkeley National Laboratory, Berkeley, CA, 94720, USA

[*]Correspondence and requests for materials should be addressed to K.-S. L. (kisuk@unist.ac.kr).

**Abstract**

Skyrmions, topologically stable magnetic solitons characterized by whirling magnetization in nanoscale magnetic elements, show promise information carriers in spintronics and spin-based quantum computing due to their unique properties—small size, stability, and controllability. In this study, we introduce a novel method of skyrmion generation through domain wall deformation dynamics. Our analytical and micromagnetic simulations demonstrate that domain wall motion exceeding the Walker threshold induces topological deformation of magnetic domain walls exhibiting Dzyaloshinskii-Moriya interaction. This deformation process catalyzes the emergence of skyrmions from magnetic domain wall structure distortion, specifically through the "Anchoring" of domain walls due to the vertical Bloch line. We elucidate the underlying mechanism of skyrmion generation, correlating it with topological transitions accompanied by burst energy dissipation through spin-wave radiation. Notably, we present robust skyrmion generation conditions through a comprehensive classification of domain wall distortion, including vertical Bloch line generation and annihilation in magnetic domain wall dynamics within a DMI system. These findings provide noble insights into topological behaviors of spin structures and offer a potential pathway for efficient, controlled skyrmion creation in the next-generation spintronic devices.

# I. Introduction

The discovery of a large interfacial Dzyaloshinskii-Moriya interaction (DMI) [1,2] in ultrathin magnetic films with perpendicular magnetic anisotropy (PMA) has received lots of academic interest as it can stabilize topological magnetic structure as skyrmions, chiral domain walls, and domain wall skyrmions.

Skyrmions are whirling magnetic spin structures that have been of academic interest due to their novel properties such as topological stability [3], tiny size [4], and ultralow threshold current density for motion [5]. These properties make skyrmions an attractive concept for various spintronic applications, including data storage [3,6], logic [7,8], neuromorphic [9-11], probabilistic [12] and quantum computing [13].

However, realizing the potential of skyrmion in real applications requires overcoming several challenging prerequisites including controllable isolated skyrmion generation/deletion, shifting to a desired position, and reliable detection. Especially, despite the predicted high stability of topologically protected skyrmions, controllable isolated skyrmion generation remains a significant challenge due to topological barriers [4,14].

Therefore, several different concepts of skyrmion generation from changing topology inside the existing magnetic domain and domain wall have been proposed, such as squeezing through geometrical constructions [15], changing from stripe domain to skyrmions [16], separating domain using local stray fields from magnetic force microscopy [17], or inducing chopping through strain/stress [18].

Furthermore, topological changes inside the magnetic domain wall are reported in two-dimensional domain wall dynamics under the influence of the DMI system in response to an external magnetic field. In one-dimensional wires, the domain wall dynamics in response to an external magnetic field is widely understood and the velocity of field-driven domain walls can be divided into two regimes due to the threshold field, the so-called Walker field. Below this field, the domain wall's velocity increases with an increasing magnetic field, whereas above it, the velocity oscillates and decreases sharply due to energy dissipation as the Walker breakdown [19-21]. However, in two-dimensional domain wall dynamics under the influence of the DMI system, the domain wall velocity has been reported as not decreasing sharply. Instead, above the Walker field, a plateau of the domain wall velocity has been observed due to energy dissipation, primarily through spin-wave emission caused by topological change of the magnetic domain wall as vertical Bloch line (VBL) generation/annihilation [21,22].

In this work, we demonstrate a new concept of skyrmion generation based on topological changes in the magnetic domain wall dynamics in a DMI system with an external magnetic field: the magnetic bubble separates from the domain structure when its dynamic distortion overs Walker field and the topological charge transition occurs accompanying with burst energy release through spin-wave emission. The separated bubble stabilized to form a skyrmion structure.

## II. Result

**Micromagnetic calculations and analysis**

We used open-source micromagnetic simulation Mumax3 software [23] to numerically solve the Landau-Lifshitz-Gilbert equation, that performs domain wall dynamics in a two-dimensional thin film with PMA and DMI, under an applied external magnetic field in the *z*-direction. The initial material parameters used in the simulation were $M_s$ = 1.13 MA/m, $A_{ex}$ = 16 pJ/m, $D$ = 3 mJ/m$^2$, $K_u$ = 1.28 MJ/m$^3$, and magnetic damping constant $\alpha$ = 0.1.

The initial state of the simulation was a domain wall at the nanostrip, separating up and down domains (as illustrated in Figure 1(a)). Because the stabilization of domain walls is affected by DMI [18], a Néel-type internal structure is introduced and becomes a Néel wall. While a steady external magnetic field is applied perpendicular to thin films, the down domain shrinks and the domain wall dynamics occur to the right side. Since we applied an external magnetic field above the Walker field, the Walker breakdown occurs and becomes a corrugated domain wall proposed by Slonczewski [24] as Figure 1(b). That corrugated domain wall contains a pair of VBLs, characterized by a 180° transition in the internal magnetization of the domain wall. The pair of VBLs in the presence of DMI becomes a pair of magnetic domain wall skyrmions [25] characterized by a 360° transition in the internal magnetization of the domain wall. Figure 1(c) shows the generation of skyrmions by domain wall dynamics after Walker breakdown under an external magnetic field. These domain wall deformation dynamics deviate from the well-known domain wall dynamics in 1D and 2D systems. Furthermore, skyrmion generations are challenging tasks, which are constrained by their topological barrier.

To confirm the change in the two-dimensional nanostrip of the magnetic structure, we analyze the topological number and topological density. The total topological number ($Q$) is given by $Q = \frac{1}{4\pi} \int q(x,y) dx dy$, where $q(x, y)$ is topological charge density with a position component $x$ and $y$ in the simulation geometry, defined by $q(x, y) = \mathbf{m} \cdot (\partial_x \mathbf{m} \times \partial_y \mathbf{m})$. In our results, we can analyze the topological change of the domain wall from the Néel wall, containing VBLs ($Q = \pm\frac{1}{2}$), domain wall skyrmions ($Q = \pm 1$), magnetic bubbles ($Q = 0$), skyrmions ($Q = -1$), and anti-skyrmions ($Q = +1$) by examining the changes in topological charge density and topological number.

**Domain wall anchoring**

We start by probing the topological charge density changes in the domain wall deformation dynamics, which is the main information necessary to understand the skyrmion generation mechanism. To achieve this, we estimated the domain wall region based on the *z*-directional magnetization condition ($|m_z| < 0.9$) with continuity that contributes to focusing only on domain wall changes. Additionally, we divided the domain wall region into two regions according to the sign of topological charge density, allowing us to examine topological changes within the domain wall. We notice that during domain wall deformation dynamics, a total topological charge of the domain wall region tends to be constrained by the conservation, observing the formation of paired spin structures with opposite topological charges into VBLs and domain wall skyrmions pairs during this process. To confirm the role of these spin structure pairs in dynamics, we defined the highest ($q_{max}$) and the lowest ($q_{min}$) positions of topological charge density and compared these positions to the overall domain wall position. While the magnitude of topological charge density at $q_{max}$ and $q_{min}$ increased by a topological transition in domain wall deformation, the mobility of those points decreased, and the positional gap between the domain wall and these points increased. This process resembles the anchoring of a ship by an anchor, or a small defect colliding and passing through high-viscosity fluid. When the domain wall anchoring persists, it can eventually exceed the domain wall stiffness, leading to domain separation, the magnitudes of topological charge density of these points steeply decrease, but the positional gap from the domain wall decreases. Because of the DMI, these separated domains can form circular domains as bubbles or skyrmions, therefore anchoring becomes the key mechanism for skyrmion generation.

**Characterizing domain wall dynamics under periodic boundary condition**

In the following, we investigated the behavior of domain walls under a $z$-directional external magnetic field with DMI, utilizing periodic boundary conditions (PBC) to prevent topological changes from the edge. We applied a $z$-directional external magnetic field $B_z$ = 110 mT. Figure 2 presents snapshots of the domain wall dynamics, the total topological number, magnetization, and topological charge density.

In this dynamics, as the external magnetic field exceeded the Walker field, the velocity of the domain wall increased steadily and transitioned into a precessional motion. During this precessional motion, a positional imbalance in the mobility led to the transition into a corrugated domain wall. In this corrugated domain wall, both the topological charge density magnitude and its positional imbalance gradually increased as domain wall anchoring, certain regions undergo a transition, forming pairs of VBLs. These VBL pairs act as anchoring sites, leading to more imbalanced domain wall velocities along the wall line with gradual denser, transitions into magnetic domain wall skyrmion pairs. Because the domain wall is continually moved by the external magnetic field, the low topological charge density magnitude regions in the domain wall 2-dimensionally expanded through anchored high topological charge density region pairs.

This expansion continued until it exceeded domain wall stiffness, ultimately leading to the separation of a bubble domain from the slightly deformed domain wall. In other words, anchoring occurred by topological changes in domain wall dynamics, the bubble domain separated from the domain wall, and the remaining domain wall would deform again.

Figure 2(a) highlights the time evolution of the total topological number $Q$, with two possible cases of topological changes during this domain wall dynamics. For both cases, the bubble domains are separated from the domain wall. Figures 2(b) and 2(c) show the different topological changes of bubbles highlighted by red and green dashed circles in Fig. 2(a) with anchoring.

The red dashed circle represents the topological number changed due to spin-wave emission during the energetically unstable antiskyrmionic magnetic structure annihilation and skyrmion generation. The green dashed circle represents the unchanged topological number, because the trivial bubble separated from the domain, subsequently shrank and annihilated.

We further investigate the generation of magnetic bubbles from domain wall dynamics, observing a reversible process of the domain wall while moving and presenting the classification of domain wall distortion. The reversible process starts with a Néel-type domain wall on the magnetic PMA thin film (Fig. 3(a)).

Applying a positive $z$-directional external magnetic field, the domain wall moves from left to right, that domain wall is transformed into a corrugated domain wall above the Walker field, and VBL pairs are created (Fig. 3(b)). Consequently, the VBL pairs transform into domain wall skyrmion pairs, with the domain wall skyrmion pairs acting as anchoring sites, the domain walls are 2-dimensionally expanded (Fig. 3(c)). These anchoring and expansions lead to the separation of bubble domains from the domain wall in two individual bubbles highlighted in blue and red boxes and the remaining domain wall highlighted in green box (Fig. 3(d)). Each separated bubble follows one of two possible topological change paths depending on skyrmion generation or not. These skyrmion generation (Case A and 1 of Fig. 3(e)) and non-generation results (Case B and 2 of Fig. 3(e)), make a total of 4 possible cases of domain wall dynamics.

Case A of the blue box involved a trivial bubble ($Q = 0$) that emitted an antiskyrmion. The antiskyrmion shrank by being energetically unfavorable due to the DMI and annihilating due to the continuum limit. This annihilating led to a change in the total topological number to $Q = -1$, accompanied by the spin-wave emission. Case B of the blue box also involved a trivial bubble ($Q = 0$) but spontaneously annihilated while maintaining $Q = 0$. Likewise, Case 1 of the red box involved a trivial bubble ($Q = 0$), transformed into one end being half skyrmion and the other end being half antiskyrmion. That half antiskyrmion was energetically unfavorable due to the DMI, became a locally dense energy area, shrinks, and annihilated while emitting a spin-wave. This annihilation led to a change in the topological number to $Q = -1$. Case 2 followed the same pattern as Case B, with the bubble spontaneously shrinking and annihilating while maintaining $Q = 0$. The remaining domain wall, highlighted in the green box, continued precession, forming new VBL pairs and repeating topological changes within the domain wall (Fig. 3(e)).

**Characterizing edge effects on domain wall dynamics**

Thus far, we discussed domain wall dynamics under the existing PBC system to prevent edge effects. However, the edge effect due to width confinement cannot be neglected in magnetic nanostrip. In the following, we focus on the domain wall dynamics within narrow, confined nanostrip width to classify the edge effects. More specifically, the width range is before overlapping two edges of magnetic domain wall deformations.

In the confined magnetic nanostrip with PMA and DMI, the edge of the domain wall becomes tilted by the DMI effect. It is well known that VBLs are generated from these tilted domain wall edges [25-27]. Also, for a sufficiently large DMI effect, such a domain wall velocity breakdown is suppressed, and the maximum velocity of the domain wall increases [21,22]. Consequently, the velocity difference between high and low topological charge density regions increases due to the edge effect, and the bubble is more likely to be separated from the domain.

Figure 4 shows the change in total topological number, magnetization, and topological charge density resulting from high DMI (such as $D = 3.0$) domain wall dynamics under a z-directional external magnetic field (e.g., $B_z$ = 100 mT) above the Walker field.

Unlike previous PBC cases, the edge effects of the domain wall during deformation dynamics show more complex changes in conventional total topological numbers. These complex changes consist of the continuous and smooth changing of total topological number as Fig. 4(a) (①~⑦), and the significant change of total topological number as Fig. 4(a) (⑦~⑨). The edges interact with the domain walls, causing both ends of the domain wall to flip more easily and potentially generating VBLs near the edges [28]. Also, while the domain walls move through the external magnetic field, VBLs change to the domain wall skyrmions while the velocity difference increases, and topological charge density becomes concentrated. This results in the domain extension along the edge and separation due to high topological density regions as Fig. 4(b) (①~⑦). The separated domains become two opposite signed topological structures, half-antiskyrmion combined in the edge, and skyrmion. The half-antiskyrmion combined in the edge that is energetically unfavored due to the DMI is condensed and annihilated into the edge. In contrast, the separated skyrmion is not annihilated and becomes stable, and significant changes in total topological numbers occur as Fig. 4(b) (⑦~⑨).

We further investigated the changes in the domain wall due to edge effects and observed significant magnetic structural changes at each edge. The domain wall deformations with edge effects can be classified as upper and lower edge deformations, as shown in Fig. 5(a), with further details provided in Fig. 5(b) and (c). The upper edge deformation can be divided into two cases, following the red boxes in Fig. 5(b). As the domain wall moves under an external magnetic field, the domain wall is tilted by the edge effect, which becomes corrugated. In both cases, the total topological number is conserved ($Q = 0$) despite the deformation occurring by the edge effect. The main difference between the left and right sides of Fig. 5(b) is whether

VBL is generated. On the left, VBL is not generated, and the tilt of the domain wall decreases, returning to the initial structure. However, on the right, VBL is generated, and the tilt of the domain wall increases, leading to domain wall detachment due to boundary interaction.

Similarly, the lower edge deformation can be classified as two cases following the blue boxes in Fig. 5(c). In both cases, the tilted domain wall generates VBL, and the corrugated structure leads to the formation of a skyrmion ($Q = -1$). The difference is that VBL or edge acts as bubble detachers while the domain wall moves. Both cases in Figure 5 (b) and (c) can be repeated under an external magnetic field, meaning that skyrmions can be generated repeatably at the edge of a confined magnetic nanostrip.

## III. Discussion

In the simulation presented, we have shown domain wall dynamics in systems with high DMI and PMA. We observed the detachment of bubbles from the domain above the Walker field, which can be stabilized into the skyrmions with topological changes.

Regarding these bubble generation mechanisms, the nanostrip width plays a significant role in determining the number of bubbles generated during domain wall dynamics. The corrugated domain walls tend to generate pairs of VBLs while conserving the total topological number. Within a confined width, there is also an edge effect making it easier for VBL generated, and a characteristic separation width between VBL formations [27,29]. Thus, at integer multiples of separation width, the number of generated VBL pairs increased. In such cases, the domain wall deformations can be spontaneously classified as shown in Fig. 3 and Fig. 5. However, at non-integer multiple of separation width, more complex deformations occur, which can be explained by overlapping of simple changes. For example, two deformations occur simultaneously, and the second deformation overlaps with the first.

As the number of generated VBLs increases, there is a corresponding increase in the number of bubble generations. Therefore, modulation of width can directly influence the number of skyrmions generated.

Based on the analytic formula from 1D domain wall motion of Slonczewski equations $q$-$\phi$ model with Bloch-like profile ansatz [18], the Euler-Lagrange equation is given by

$$\frac{\partial \phi}{\partial t} + \frac{\alpha}{\Delta}\frac{\partial q}{\partial t} = \gamma_0 H$$

At the Walker field $H_w$, domain wall dynamics is steady state threshold, $\dot{\phi} = 0$, then the critical velocity named Walker velocity $v_w$ can be driven as

$$v_w = \gamma_0 \frac{\Delta}{\alpha} H_w \approx \frac{\pi}{2}\gamma_0 \Delta H_{\text{DMI}} = \frac{\pi}{2}\gamma \frac{D}{M_s}$$

With $\gamma_0 = \mu_0 \gamma$ where $\gamma$ is the gyromagnetic ratio, $\Delta = \sqrt{A_{\text{ex}}/K_{\text{eff}}}$ is the domain wall width parameter and the DMI field $\mu_0 H_{\text{DMI}} = \frac{D}{M_s \Delta}$ [18]. This Walker velocity depends on intrinsic parameters such as saturation magnetization, anisotropy constant, damping constant, and DMI constant. Especially in the DMI case, the domain wall velocity at the Walker field is increased linearly with DMI strength and domain wall width parameter. However, the suggested domain wall deformation mechanism has a limited range of DMI strength. In the low DMI strength cases, the difference of domain wall velocity with the anchoring site is lower, VBLs preferred to disperse into domain wall, and it is difficult for these VBLs to change to bubbles. In high DMI strength cases, the domain walls tend to become maze structures.

We discuss the position of topological structure creation inside the domain wall, which is generated sideways away from domain wall expansion trajectories. The transition of the topological structure inside the magnetic domain wall changes into Néel-type domain wall, corrugated domain wall, and VBLs. At that time, VBL becomes smaller due to the influence of DMI and external magnetic field and changes to the form of domain wall skyrmion with $Q = \pm1$. Recently, the skyrmion Hall effect was observed in domain wall skyrmion under an external magnetic field [30]. In our results, VBL pairs with opposite skyrmion numbers transform into domain wall skyrmion pairs, this process induces skyrmion Hall effects in opposite *y*-directions along the same domain wall. This skyrmion Hall effect, combined with the domain wall anchoring effect, also plays a decisive role in the separation of the domain, as shown in the blue box in Fig.3. Thus, the skyrmion Hall effect on VBL and domain wall skyrmion pairs is also crucial to skyrmion generation.

We finally discuss the domain wall dynamics above the Walker field and skyrmion generation supported by energy considerations. For a high external magnetic field above the Walker field, the 2-dimensional domain wall velocity follows the plateau. The domain wall magnetization becomes more complex deformations due to their energy dissipation by fast precessional motion, the creation and annihilation of magnetic solitons. Also, the external magnetic field gives a steady increment of energy to the system until energy dissipation occurs, these creation and annihilation processes are periodically induced.

These relations between precession and the external field have been shown based on the momentum conservation scheme by the spatial and temporal average of the Slonczewski equation [22,24].

$$\ll \frac{\partial \phi}{\partial t} \gg + \frac{\alpha}{\Delta} \ll \frac{\partial q}{\partial t} \gg = \gamma_0 H$$

We can derive domain wall magnetization precession under the assumption that domain wall velocity is fixed as Walker velocity. [22]

$$\ll \frac{\partial \phi}{\partial t} \gg \approx \gamma_0 (H - H_\text{w}).$$

The domain wall magnetization precession is intimately related to the creation and annihilation of pairs of VBLs. In our case, the formation of magnetic structures in the domain wall is paired through tends to conserve of total topological number in the domain wall. In that pair, the positive topological charges are annihilated because energetically unstable due to interfacial DMI. Therefore, analysis of the total number of positive topological charge densities changing can refer to the frequency of creation and annihilation of pairs of VBLs. By analysis of positive topological charge densities changing frequency $f_{\text{VBL+}}$ by FFT, one gets $f_{\text{VBL+}} \approx \frac{\gamma_0}{N_{\text{VBL}}}(H_\text{a} - H_\text{w})$. This means one-time domain wall magnetization precession occurs, the $N_{\text{VBL}}$ number of positively charged VBLs are created, and that frequency is related to the magnitude of the external magnetic field. Since the creation of VBLs occurs in pairs and leads to the separation of bubble domains from the domain wall, linked to our skyrmion generation mechanism. Therefore, we can modulate skyrmion generation frequencies by adjusting an external magnetic field.

## IV. Conclusion

We have explored a novel concept for generating skyrmions through topological changes in the dynamics of magnetic domain walls above the Walker field, in 2D magnetic thin film with DMI under the influence of an external magnetic field.

We observed that under the influence of an external magnetic field exceeding the Walker field, domain walls undergo a transition into corrugated domain walls, accompanied by the generation of VBLs. These VBLs act as anchoring sites and are transformed into the domain wall skyrmions, affecting the domain wall's velocity by anchoring and leading to local topological charge density changes as detachment of magnetic bubbles from domain and classified into types of skyrmion generation mechanisms. These topological changes can be analyzed by the domain wall dynamics to energy dissipation and changing of topological charge density profiles. Also, depending on the nanostrip edges, widths, and material parameters such as saturation magnetization, anisotropy constant, damping constant, and DMI constant. These changes occur periodically above the Walker field, the creating topological structure positions are changed due to the skyrmion Hall effect of the domain wall skyrmions, and the topological density changing frequency depends on the magnitude of the external magnetic field. Based on such correlation, our study offers valuable insight into the controllable creation of skyrmions through the domain wall dynamics.

Finally, our topological changes analysis provides a deeper understanding and links between various spin structures such as domain walls, VBL, domain wall skyrmion, and skyrmions that have been studied from the past to the present. Further exploration of skyrmion generation mechanisms and their optimization could pave the way for practical skyrmion-based devices and future works for domain wall-switching applications.

# References.

[1] I. Dzyaloshinsky, J Phys Chem Solids **4**, 241 (1958).
[2] T. Moriya, Physical Review **120**, 91 (1960).
[3] J. Sampaio, V. Cros, S. Rohart, A. Thiaville, and A. Fert, Nature Nanotechnology **8**, 839 (2013).
[4] N. Nagaosa and Y. Tokura, Nature Nanotechnology **8**, 899 (2013).
[5] A. Fert, N. Reyren, and V. Cros, Nature Reviews Materials **2** (2017).
[6] A. Fert, V. Cros, and J. Sampaio, Nature Nanotechnology **8**, 152 (2013).
[7] S. J. Luo, M. Song, X. Li, Y. Zhang, J. M. Hong, X. F. Yang, X. C. Zou, N. Xu, and L. You, Nano Letters **18**, 1180 (2018).
[8] D. H. Jung, H. S. Han, N. Kim, G. Kim, S. Jeong, S. Lee, M. Kang, M. Y. Im, and K. S. Lee, Physical Review B **104**, L060408 (2021).
[9] J. Grollier, D. Querlioz, K. Y. Camsari, K. Everschor-Sitte, S. Fukami, and M. D. Stiles, Nature Electronics **3**, 360 (2020).
[10] K. M. Song *et al.*, Nature Electronics **3**, 148 (2020).
[11] Z. Y. Yu *et al.*, Nanoscale Advances **2**, 1309 (2020).
[12] D. Pinna, F. A. Araujo, J. V. Kim, V. Cros, D. Querlioz, P. Bessiere, J. Droulez, and J. Grollier, Physical Review Applied **9** (2018).
[13] Y. Ji *et al.*, Adv Mater **36**, e2312013 (2024).
[14] W. Jiang *et al.*, Science **349**, 283 (2015).
[15] S. Woo *et al.*, Nature Materials **15**, 501 (2016).
[16] A. Casiraghi, H. Corte-Leon, M. Vafaee, F. Garcia-Sanchez, G. Durin, M. Pasquale, G. Jakob, M. Klaui, and O. Kazakova, Communications Physics **2** (2019).
[17] F. Büttner *et al.*, Nature Nanotechnology **12**, 1040 (2017).
[18] A. Thiaville, S. Rohart, É. Jué, V. Cros, and A. Fert, EPL (Europhysics Letters) **100** (2012).
[19] A. Dourlat, V. Jeudy, A. Lemaitre, and C. Gourdon, Physical Review B **78** (2008).
[20] P. J. Metaxas, J. P. Jamet, A. Mougin, M. Cormier, J. Ferré, V. Baltz, B. Rodmacq, B. Dieny, and R. L. Stamps, Physical Review Letters **99** (2007).
[21] Y. Yoshimura *et al.*, Nature Physics **12**, 157 (2015).
[22] V. Krizakova, J. P. Garcia, J. Vogel, N. Rougemaille, D. D. Chaves, S. Pizzini, and A. Thiaville, Physical Review B **100** (2019).
[23] A. Vansteenkiste, J. Leliaert, M. Dvornik, M. Helsen, F. Garcia-Sanchez, and B. Van Waeyenberge, AIP Advances **4** (2014).
[24] J. C. Slonczewski, C. D. Graham, and J. J. Rhyne, 1972), pp. 170.
[25] R. Cheng *et al.*, Physical Review B **99** (2019).
[26] T. Herranen and L. Laurson, Physical Review B **92** (2015).
[27] T. Herranen and L. Laurson, Physical Review B **96** (2017).
[28] K. Yamada and Y. Nakatani, Applied Physics Express **8** (2015).
[29] J. Slonczewski, Journal of Magnetism and Magnetic Materials **12**, 108 (1979).

[30] S. U. Han, W. Kim, S. K. Kim, and S.-G. Je, Physical Review B **109** (2024).

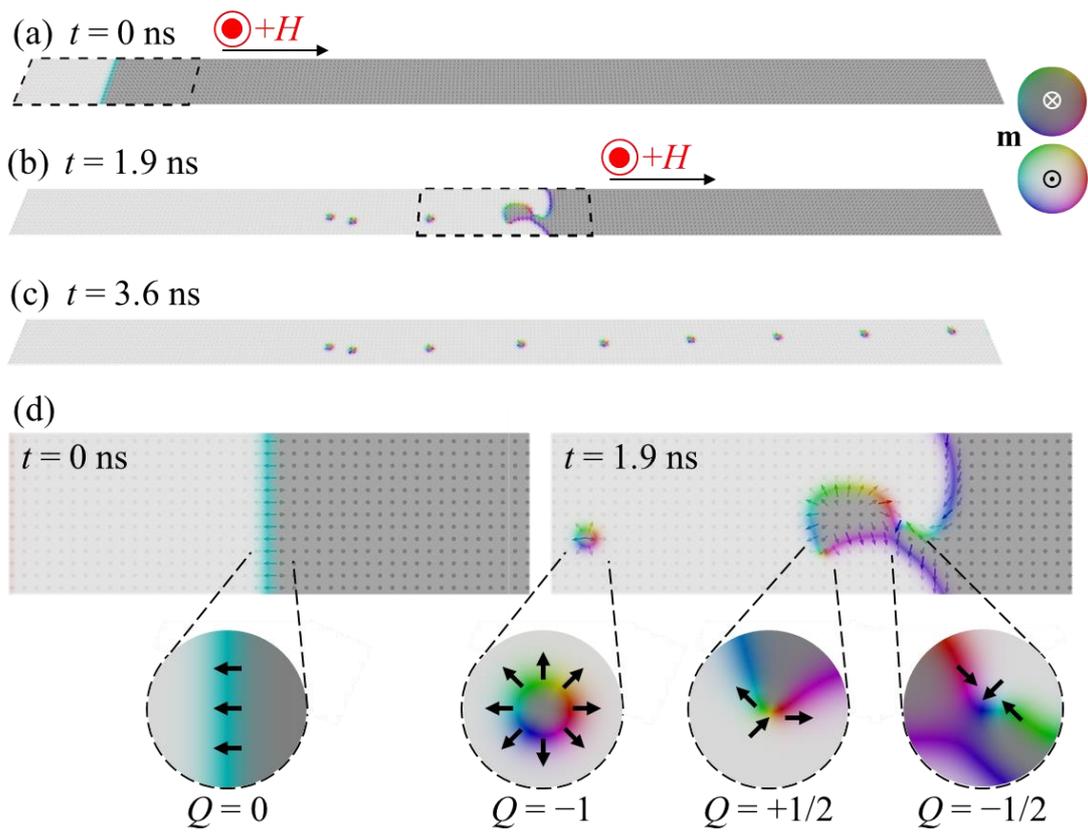

Figure 1. An example of domain wall changes by the *z*-directional external magnetic field using micromagnetic simulations. (a) initial state (*t* = 0) is the domain wall separate the up and down domain. (b) The domain wall moved by external field *H* higher than Walker field, becomes a corrugated wall. (c) The domain wall reaches the end of the nanostrip, while generation skyrmions. (d) The frame region of the moving window of (a) and (b).

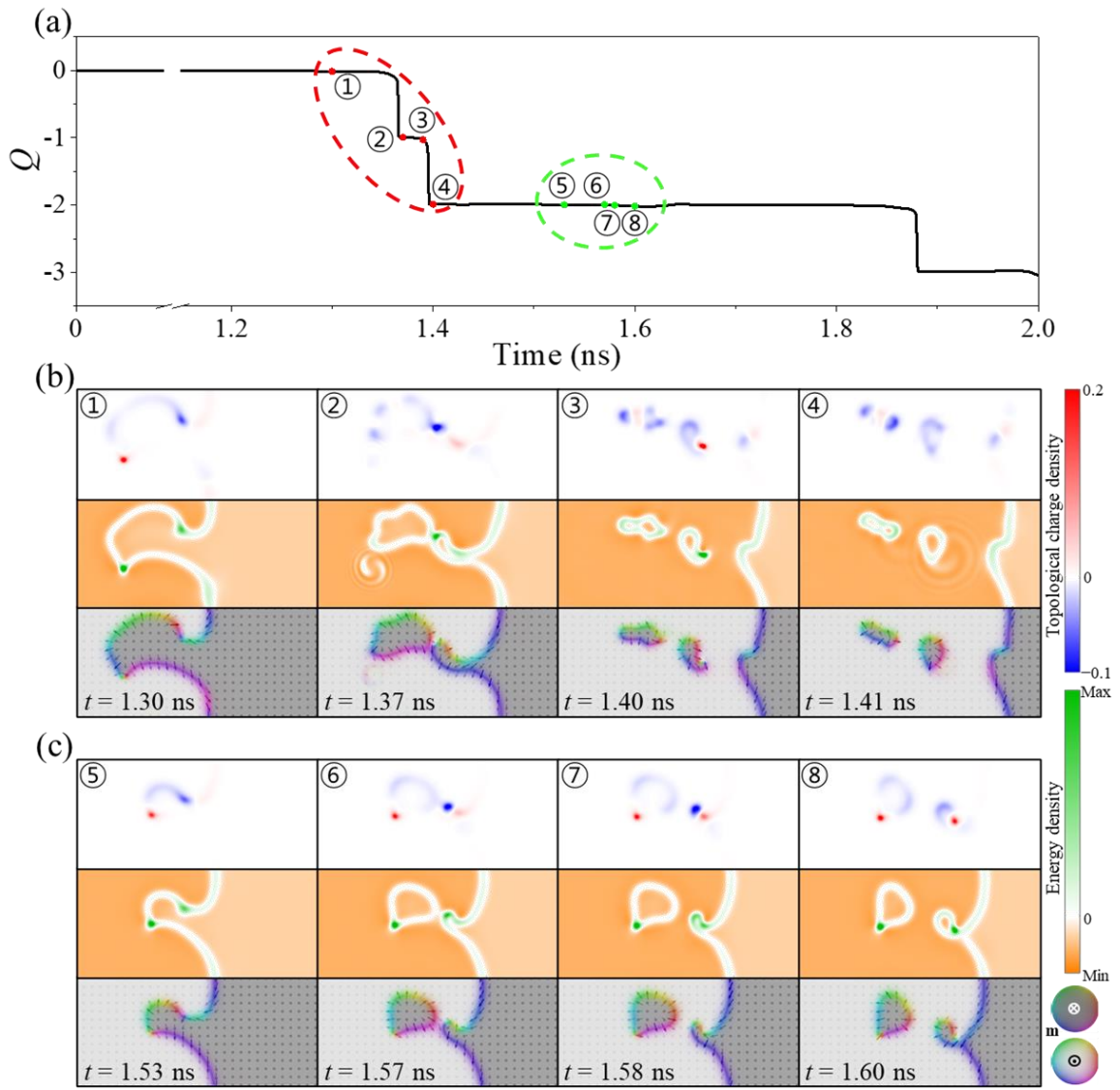

Figure 2. Simulation results of domain wall dynamics in the ultrathin film under *z*-directional external magnetic field with DMI, PMA, and using PBC. (a) Time evolution of total topological number *Q*. (b) and (c) The snapshots of topological charge density, total energy density, and magnetization configuration at the red dashed circle and green dashed circle.

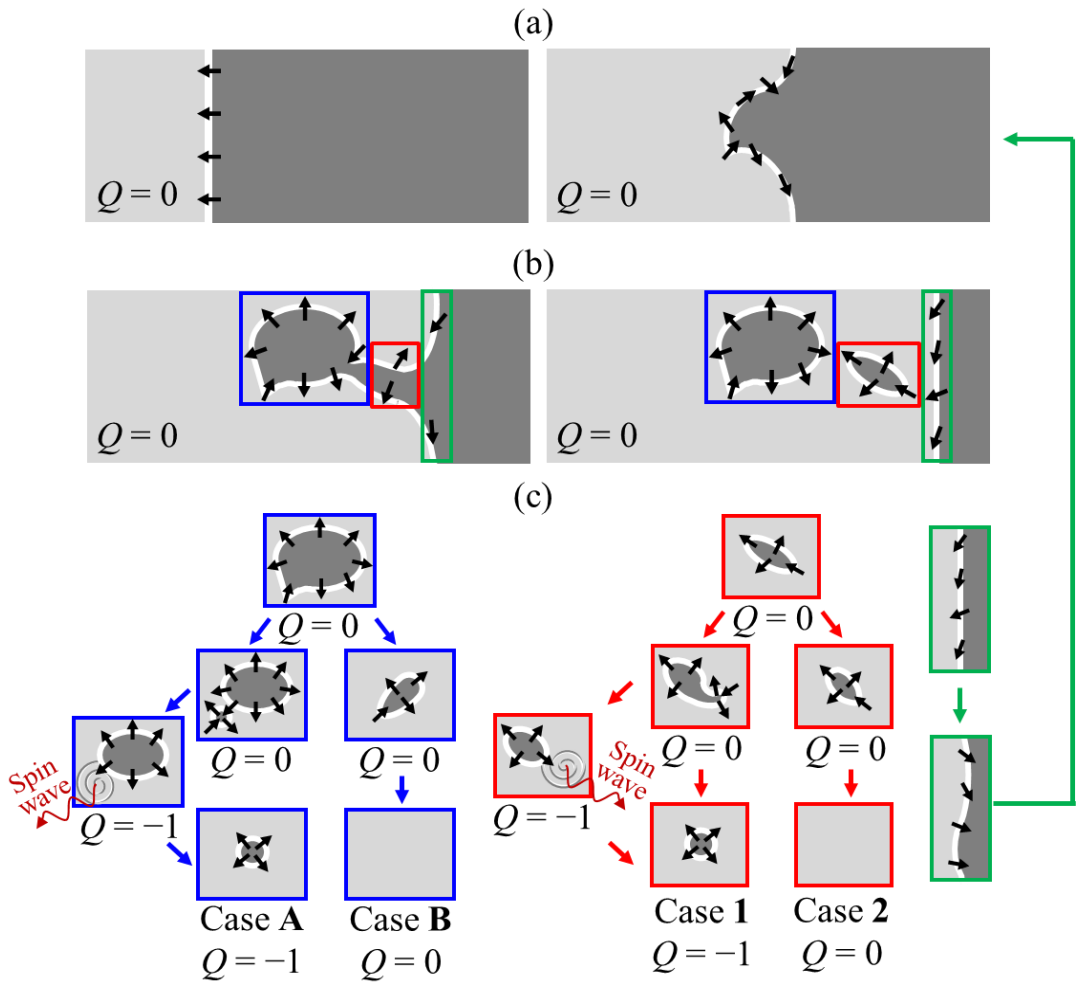

Figure 3. (a)-(f) Schematic of the domain wall configuration changes and topological number changes using PBC. Starting from (a) Néel-type domain wall, (b) changing to the corrugated wall and VBL pair generation, (c) VBL pair transition to domain wall skyrmion pair, (d) anchoring of domain wall due to VBL, (e) two bubbles generations repeatably occurs due to domain wall dynamics. (f) Depending on the topological changes of the two bubble cases, the generation of skyrmion is determined.

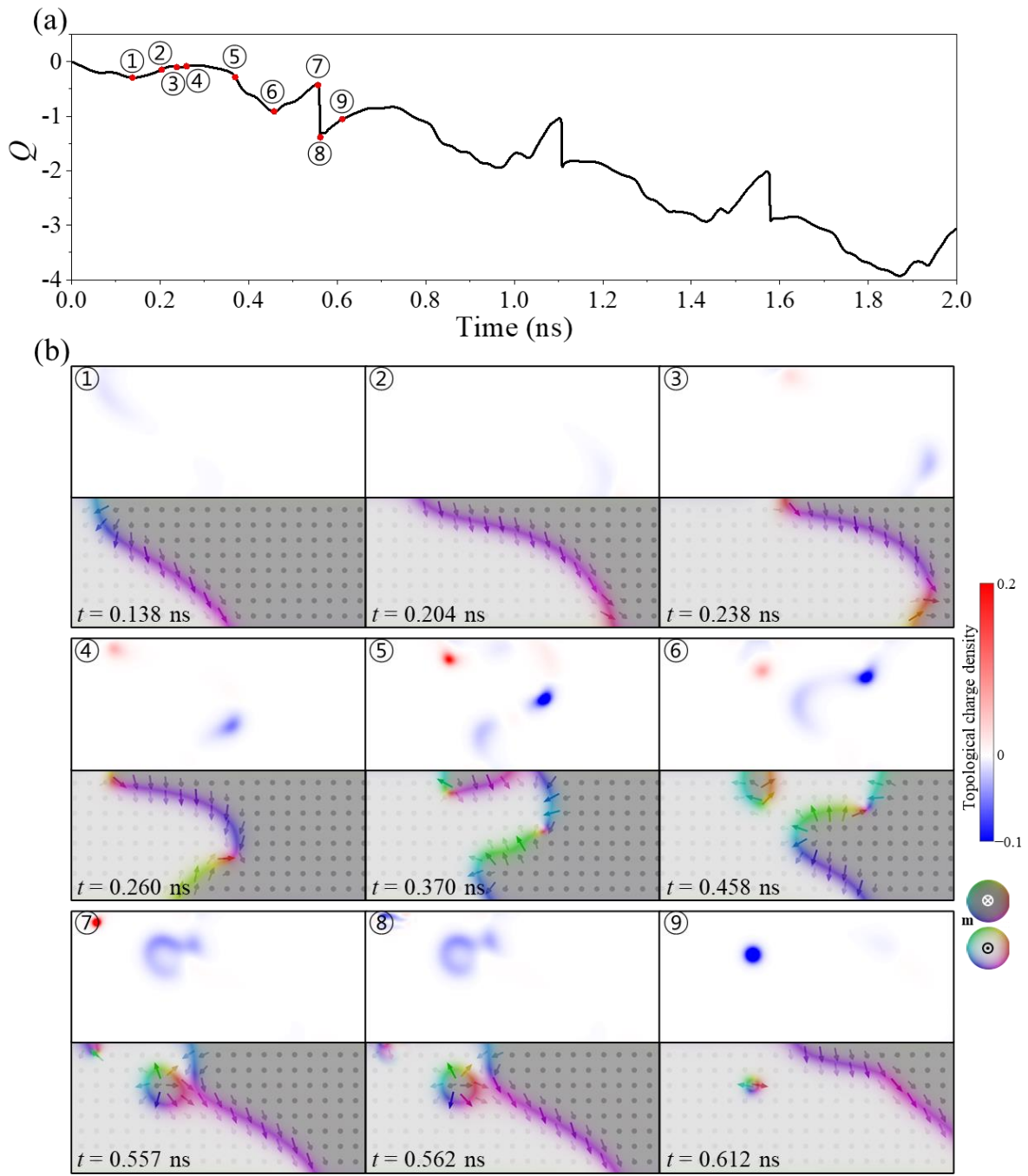

Figure 4. Simulation results of domain wall dynamics in the ultrathin film under a $z$-directional external magnetic field with DMI, PMA, and width confinement. (a) Time evolution of total topological number $Q$. (b) The snapshots of topological charge density, and magnetization configuration following numbers in Fig. 4(a).

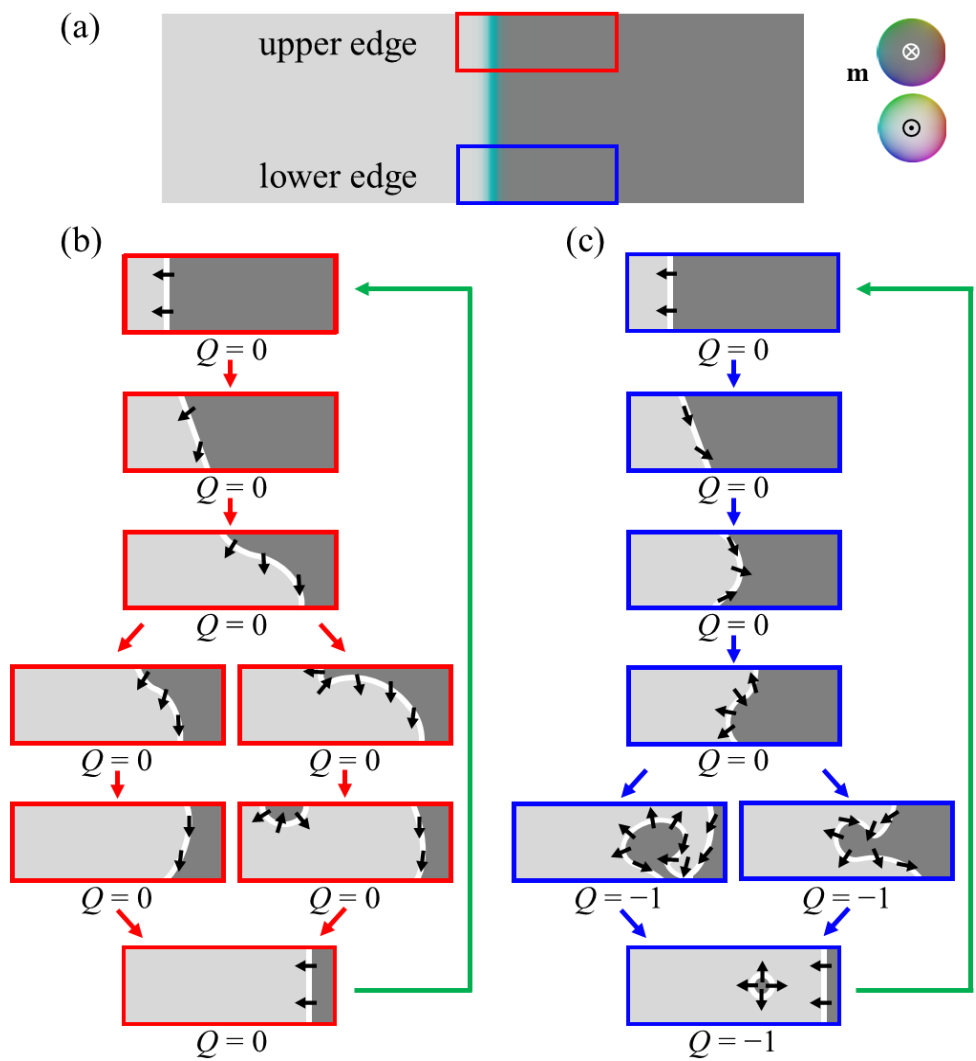

Figure 5. (a)-(c) Schematic of the domain wall configuration changes and topological number changes with width confinement. The edge effects on domain wall configuration can be classified by upside edge deformations and downside edge deformation because of tilted angle and VBL motion differences.